\documentclass[prl,reprint,showpacs,amsmath,amssymb,superscriptaddress]{revtex4-1}
\usepackage{bm}
\usepackage[pdftex]{graphicx,hyperref}
\hypersetup{colorlinks=true, anchorcolor=blue, linkcolor=blue, citecolor=blue, filecolor=blue, urlcolor=blue}

\begin{document}
\title{Three-Dimensional Coupled Dynamics of Two-Fluid Model in Superfluid $^4$He: Deformed Velocity Profile of Normal Fluid in Thermal Counterflow}
\author{Satoshi Yui}
\affiliation{Department of Physics, Osaka City University, 3-3-138 Sugimoto, Sumiyoshi-ku, Osaka 558-8585, Japan}
\author{Makoto Tsubota}
\affiliation{Department of Physics, Osaka City University, 3-3-138 Sugimoto, Sumiyoshi-ku, Osaka 558-8585, Japan}
\affiliation{The OCU Advanced Research Institute for Natural Science and Technology (OCARINA), Osaka City University, 3-3-138 Sugimoto, Sumiyoshi-ku, Osaka 558-8585, Japan}
\author{Hiromichi Kobayashi}
\affiliation{Department of Physics \& Research and Education Center for Natural Sciences, Hiyoshi Campus, Keio University, 4-1-1 Hiyoshi, Kohoku-ku, Yokohama 223-8521, Japan}
\date{\today}

\begin{abstract}
The coupled dynamics of the two-fluid model of superfluid $^4$He is numerically studied for quantum turbulence of the thermal counterflow in a square channel.
We combine the vortex filament model of the superfluid and the Navier--Stokes equations of normal fluid.
Simulations of the coupled dynamics show that the velocity profile of the normal fluid is deformed significantly by superfluid turbulence as the vortices become dense.
This result is consistent with recently performed visualization experiments.
We introduce a dimensionless parameter that characterizes the deformation of the velocity profile.
\end{abstract}
\maketitle

{\it Introduction.}---When a fluid system consists of several continuous fields, their interaction as well as the generation mechanism of ordered or disordered states is complex.
Such complex systems are ubiquitous in nature \cite{davidson15}, for example, in magnetohydrodynamics \cite{moreau98,kobayashi08} and two-component Bose--Einstein condensates of cold atomic gases \cite{takeuchi10}.  The same result is found for the two-fluid model in low-temperature physics.
Since the proposal by Tisza \cite{tisza38} and Landau \cite{landau41}, the two-fluid model has functioned as a powerful phenomenological model in superfluidity and superconductivity \cite{tilley90}.
The model states that the system consists of an inviscid superfluid (density $\rho_\text{s}$) and a viscous normal fluid (density $\rho_\text{n}$) with two velocity fields, ${\bm v}_\text{s}$ and ${\bm v}_\text{n}$.
A superfluid is subject to the severe quantum-mechanical constraint, and any rotational motion is sustained only by quantized vortices with quantum circulation $\kappa$.
The two-fluid model has been useful for understanding various phenomena in low-temperature physics \cite{tilley90}, but its coupled dynamics has seldom been studied.

The system most characteristic of the two-fluid model is thermal counterflow in superfluid $^4$He, which exhibits a typical stage of quantum hydrodynamics and quantum turbulence \cite{vinen02,halperin09,tsubota13,barenghi14,tsubota17}.
In usual experiments of thermal counterflow, superfluid $^4$He is confined within a channel, with one closed end and the other end connected to a helium bath.
Upon heating the closed end, the normal fluid flows towards the helium bath, and the superfluid flows in the opposite direction to satisfy the total mass conservation $\int (\rho_\text{n} {\bm v}_\text{n} + \rho_\text{s}{\bm v}_\text{s}) \text{d}{\bm r}={\bm 0}$, with the integral performed over the cross-section of the channel.
When the relative velocity $|{\bm v}_\text{ns}| = |{\bm v}_\text{n} - {\bm v}_\text{s}|$ between two fluids exceeds the critical value, the superfluid becomes turbulent, and consists of a tangle of quantized vortices \cite{vinen57a,vinen57b,vinen57c,vinen58,feynman55}.
The vortex filament model (VFM) is suitable for superfluid $^4$He;
the numerical simulations of the VFM in a bulk under the prescribed normal fluid flow revealed many properties of thermal counterflow \cite{schwarz88,adachi10}.

Two kinds of experiments in counterflow require theoretical and numerical approaches beyond the present state.
One is the recent visualization experiments \cite{guo14}.
Marakov {\it et al.} observed the flow profiles of the normal fluid in a square channel \cite{marakov15}.
By increasing the heat, the profile of the normal fluid velocity changed from the laminar Poiseuille, via the laminar tail-flattened, and eventually to the turbulent flow.
To date, the laminar tail-flattened profile, in which the tail part becomes flattened, has not been reported in classical hydrodynamics.
The other is the experiments observing several kinds of superfluid turbulence \cite{tough82}.
When the aspect ratio of the cross-section of the channel was low, the system showed two turbulent states, namely T1 and T2.
If the aspect ratio of the channel was high, the counterflow exhibited only a single turbulent state T3.
There are little information of these different turbulent states and their dependence on the aspect ratio.

Based on the experiments, we consider two important effects from theoretical and numerical perspectives.
First, we consider the boundary effects of the channel walls.
A few studies simulated the VFM under the prescribed normal fluid-velocity profile between two parallel plates or in a square channel to find inhomogeneous vortex tangles affected by the boundaries \cite{baggaley13,baggaley15,yui15a,yui15b}.
The other is the coupled dynamics of superfluid and normal fluid.
Because the VFM is Lagrangian and the Navier--Stokes equations describing the normal fluid are Eulerian, it is difficult to coordinate the two different schemes.
There have been only limited numerical works on the coupled dynamics, which is confined to the case of a bulk \cite{kivotides00,kivotides07}.

A few theoretical and numerical studies are useful in clarifying these observations.
Melotte and Barenghi performed the linear stability analysis of normal fluid affected by a vortex tangle through mutual friction to study the transition from T1 to T2 \cite{melotte98}.
Khomenko {\it et al.} studied numerically the three-dimensional (3D) coupled dynamics of the two-fluid model between two parallel plates \cite{khomenko16}.
They simplified the Navier--Stokes equations of the normal fluid by spatially averaging ${\bm v}_\text{n}$ over the two directions parallel to the plates.
The profile of the normal fluid was deformed, which differed from the observations \cite{marakov15}.
The numerical studies of two-dimensional counterflow were performed \cite{galantucci15}, which differs from the 3D system.
Saluto {\it et al.} studied analytically the velocity profile of the normal fluid by using the one-fluid model \cite{saluto14,saluto15}, and found that the velocity profile could be flattened by superfluid turbulence.

{\it Coupled dynamics of two fluids.}---We used the VFM for quantized vortices and the Navier--Stokes equations for normal fluid, then connected them by mutual friction \cite{kivotides00,kivotides07}.
The VFM represents a quantized vortex as a parametric form as ${\bm s} = {\bm s}(\xi,t)$ \cite{schwarz85}.
At zero temperature, the velocity $\dot{\bm s}_0 (\xi,t)$ on the filaments is given by $\dot{\bm s}_0 = {\bm v}_{\text{s},\omega} + {\bm v}_\text{s,b} + {\bm v}_\text{s,a}$, where ${\bm v}_{\text{s},\omega}$ is the velocity field produced by all the vortex filaments, ${\bm v}_\text{s,b}$ is the velocity field produced by solid boundaries, and ${\bm v}_\text{s,a}$ is the applied uniform flow of the superfluid.
The velocity field ${\bm v}_{\text{s},\omega}$ is given by the Biot--Savart law ${\bm v} _{\text{s},\omega} ({\bm x}) = (\kappa/4 \pi) \int _{\cal L} ({\bm s} _{1} - {\bm x}) \times \text{d} {\bm s} _{1} / |{\bm s} _{1} - {\bm x} |^{3}$, where ${\bm s}_1$ refers to a point on the filament, and the integration is performed along all of the filaments \cite{adachi10}.
At finite temperatures, the quantized vortices are affected by the normal fluid through the mutual friction force, and the velocity $\dot{\bm s}$ of quantized vortices at a point ${\bm s}$ is given by
\begin{equation}
  {\dot{\bm s}} = \dot{\bm s}_0 + \alpha {\bm s}' \times ({\bm v}_\text{n} - \dot{\bm s}_0) - \alpha' {\bm s}' \times [{\bm s}' \times ({\bm v}_\text{n} - \dot{\bm s}_0)],
  \label{vortex}
\end{equation}
where $\alpha$ and $\alpha'$ are the temperature-dependent coefficients of the mutual friction, and ${\bm s}'$ is the unit vector along the filaments.

The dynamics of the normal fluid obeys the Navier--Stokes equations \cite{donnelly91},
\begin{equation}
  \rho_\text{n} \left[ \frac{ \partial {\bm v}_\text{n} }{\partial t} + ({\bm v}_\text{n} \cdot \nabla) {\bm v}_\text{n} \right] = -\frac{\rho_\text{n}}{\rho} \nabla p + \eta_\text{n} \nabla^2 {\bm v}_\text{n} + {\bm F}_\text{ns},
  \label{eq:navier}
\end{equation}
where ${\nabla p}/\rho = {\nabla P}/\rho + \rho_\text{s} S \nabla T / \rho_\text{n}$ is the sum of the pressure gradient and temperature gradient, $\eta_\text{n}$ is the viscosity of normal fluid, and ${\bm F}_\text{ns}$ is the mutual friction force per unit volume.
We close these equations using the incompressible condition $\nabla \cdot {\bm v}_\text{n} = {\bm 0}$ for Eq. (\ref{eq:navier}).
In this study, we focus on the deformed profile of normal fluid velocity, so we employ the coarse-grained expression of the mutual friction force: ${\bm F}_\text{ns} ({\bm x}) = (1/\Omega' ({\bm x})) \int_{ {\mathcal L}' ({\bm x})} {\bm f} (\xi)\text{d}\xi$, where ${\bm f}(\xi)/\rho_\text{s} \kappa = \alpha {\bm s}' \times [ {\bm s}' \times ( {\bm v}_\text{n} - \dot{\bm s}_0 ) ] + \alpha ' {\bm s}' \times ({\bm v}_\text{n} - \dot{\bm s}_0)$ \cite{donnelly91}.
Here, $\Omega '({\bm x})$ is a local sub volume at ${\bm x}$, and the integral path ${\mathcal L}' ({\bm x})$ represents vortex lines in the sub volume $\Omega '({\bm x})$.
Details are described in Section I of the supplemental material.
Because ${\bm v}_\mathrm{n}$ is discretized in this formulation, we use a linear interpolation to obtain the value of ${\bm v}_\mathrm{n}$ at the position ${\bm s}$ in Eq. (\ref{vortex}).

Normal fluid has two dimensionless parameters: one is a Reynolds number $Re = \rho_\text{n} \overline{v}_\text{n} D / \eta_\text{n}$, and the other is a parameter $\phi$ characterizing the mutual friction.
Here, $v_\text{n}$ is the streamwise component of ${\bm v}_\text{n}$, $\overline{v}_\text{n}$ is the averaged value of $v_\text{n}$ over the channel cross-section, and $D$ is the half width of the cross-section.
The parameter $\phi$ is introduced as follows.
When the mutual friction is negligible, Eq. (\ref{eq:navier}) becomes the usual Navier--Stokes equations, and the normal fluid velocity should take the Poiseuille profile below the critical Reynolds number.
If the mutual friction becomes larger than the viscous forces, it can change the velocity profile from the Poiseuille profile.
Their competition is characterized by a ratio $|{\bm F}_\text{ns}|/| \eta_\text{n} \nabla^2 {\bm v}_\text{n}|$.
In a thermal counterflow, the counterflow condition $\overline{\bm v}_\text{ns} = \overline{\bm v}_\text{n} - \overline{\bm v}_\text{s} = \rho \overline{\bm v}_\text{n} / \rho_\text{s}$ is satisfied, and we can use the Gorter--Mellink relation $\overline{\bm F}_\text{ns} = g \rho_\text{s} \kappa \alpha L \overline{\bm v}_\text{ns}$ \cite{tough82,donnelly91}.
Here, $L$ is a vortex line density: $L = (1/\Omega) \int_{\mathcal L} \text{d} \xi$ with a sample volume $\Omega$.
Then, the ratio is reduced to
\begin{equation}
  \phi = \frac{g \rho \kappa \alpha L D^2}{\eta_\text{n}}.
  \label{eq:phi}
\end{equation}
To calculate the value of Eq. (\ref{eq:phi}), we assume isotropic tangle: $g=2/3$.
A similar parameter is also introduced in Ref. \cite{melotte98}.

{\it Numerical simulation.}---Numerical simulations of both fluids were performed under the following conditions.
The computing box is $1.0 \times 1.0 \times 1.0 ~\mathrm{cm^3}$.
The periodic boundary condition is applied in the $x$ direction, whereas the solid boundary condition is applied in the $y$ and $z$ directions.
The time resolutions of both fluids are $\Delta t = 1.0 \times 10 ^{-3} ~\mathrm{s}$.
The vortex lines are discretized into a number of points held at a minimum-space resolution of $\Delta \xi = 8.0 \times 10 ^{-3} ~\mathrm{cm}$.
We reconnected two vortices artificially when the vortices approached each other more closely than $\Delta \xi$.
We eliminated vortices that were shorter than $5 \times \Delta \xi$.
The integration of time for the normal fluid is achieved using the second-order Adams--Bashforth method, and the second-order finite-difference method was adopted for spatial differentiation.
The inhomogeneous spatial grid of the normal fluid is $N_x \times N_y \times N_z = 15^3$.
The temperature is $T = 2.0 ~\mathrm{K}$, and the values of the various parameters are shown in Ref. \cite{barenghi83}.
The $y$ and $z$ components of ${\bm F}_\text{ns}$ are neglected while calculating Eq. (\ref{eq:navier}), and $F_\text{ns}^x$ is averaged over the $x$-direction.
We regard the counterflow condition as ${\bm v}_\text{s,a} = - \rho_\text{n} \overline{\bm v}_\text{n} / \rho_\text{s}$ \cite{yui15a}.
The mean velocity of the normal fluid is a constant parameter.
The numerical simulation begins with eight randomly oriented loops for quantized vortices and the Poiseuille flow for a normal fluid.

\begin{figure}
  \centering
  \includegraphics[width=1.0\linewidth]{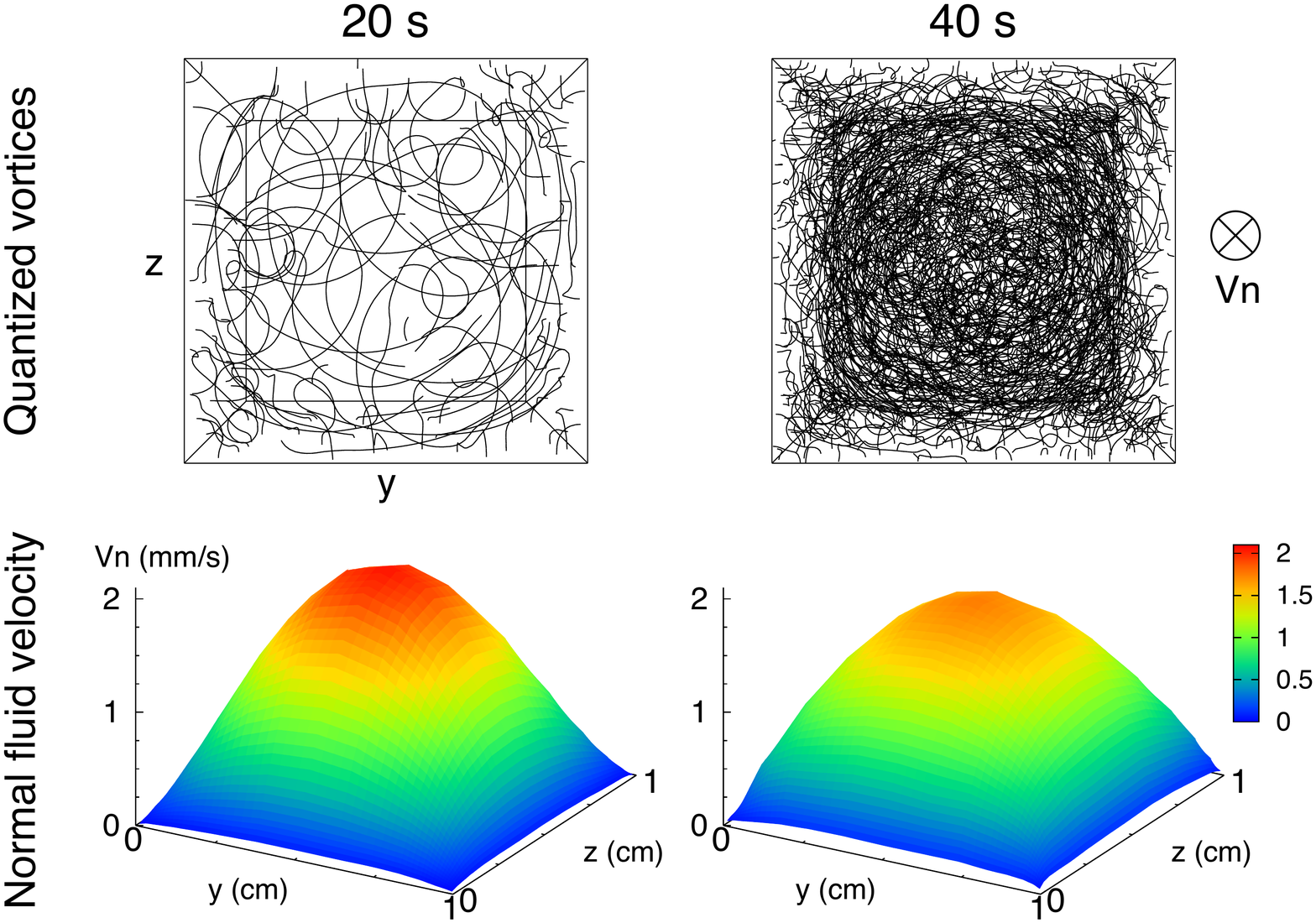}
  \caption
  {
  The quantized vortices and normal fluid velocity profile in thermal counterflow are obtained by numerical simulation of the coupled dynamics at $\overline{v}_\text{n} = 1.0 ~\mathrm{mm/s}$ and $T=2.0 ~\text{K}$ ($Re=300$).
  ({\it top})
  The black curves show vortex filaments that represent quantized vortices.
  ({\it bottom})
  The color map shows the $x$-component of the normal fluid velocity.
  Here, the mean velocity $\overline{v}_\text{n}$ of the normal fluid is constant.
  \label{ss_vn_profiles.pdf}
  }
\end{figure}

Figure \ref{ss_vn_profiles.pdf} shows the quantized vortices and normal fluid velocity profile for $\overline{v}_\text{n} = 1.0 ~\mathrm{mm/s}$
\footnote
{
The typical dynamics can be seen in the movie in the supplemental material.
}.
At $t=20 ~\mathrm{s}$, the vortices are not so dense that the normal fluid still keeps the Poiseuille profile.
However, when the vortices become dense at $t=40 ~\mathrm{s}$, the strong mutual friction significantly deforms the normal fluid velocity profile; the central velocity decreases much from that of the Poiseuille profile; in contrast, the velocity near the channel walls increases.
Thus, the velocity profile of the normal fluid may be deformed when the vortices are sufficiently dense, and when the parameter $\phi ~( \propto L)$ of Eq. (\ref{eq:phi}) is large.

\begin{figure}
  \centering
  \includegraphics[width=1.0\linewidth]{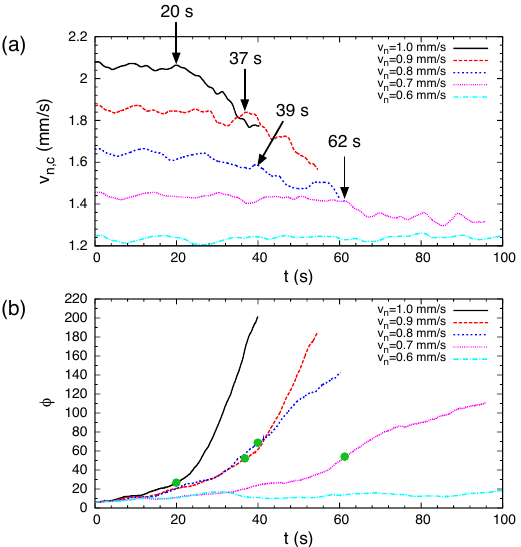}
  \caption
  {
  (a)
  The central velocity $v_\text{n,c}$ of a normal fluid as a function of time at different mean velocity $\overline{v}_\text{n}$ of the normal fluid.
  The values fluctuate around the constant values in the early stage, and decrease after the onset time shown by the arrows.
  (b)
  Mutual friction force $\phi$ of Eq. (\ref{eq:phi}) as a function of time.
  The values significantly increase after the deformation begins, which is marked by the green circles.
  \label{uc_phi.pdf}
  }
\end{figure}

The deformation of the normal fluid velocity profile can be measured by the reduction of the normal fluid velocity $v_\text{n,c}$ at the center $(y,z)=(0.5 ~\mathrm{cm}, 0.5~\mathrm{cm})$ of the channel cross-section.
By increasing the mean relative velocity $\overline{v}_\text{ns,a} = \overline{v}_\text{n} - v_\text{s,a} = \rho \overline{v}_\text{n}/\rho_\text{s}$, the eventual vortex line density increases according to the stationary solution of Vinen's equation \cite{vinen57c,tough82,yui15a}.
Then, the eventual value of $\phi$ increases with $\overline{v}_\text{n}$, and we can expect the significant deformation of the normal fluid velocity profile at larger $\overline{v}_\text{n}$.
As shown in Fig. \ref{uc_phi.pdf}(a), the significant reduction of $v_\text{n,c}$ is observed for the large mean velocity $\overline{v}_\text{n}=1.0, ~0.9, ~0.8 ~\mathrm{mm/s}$.
When the vortices become too dense, we cannot continue the numerical simulation, and we stop it.
At $\overline{v}_\text{n} = 0.7 ~\mathrm{mm/s}$, the value of $v_\text{n,c}$ starts to decrease at $t \approx 60 ~\mathrm{s}$, and it fluctuates after $t \approx 65 ~\mathrm{s}$ around some constant value that is different from the initial one
\footnote
{
As shown in Section II of the supplemental material, the value of $v_\text{n,c}$ briefly decreases, but recovers to about the initial value at $\overline{v}_\text{n}=0.7 ~\mathrm{mm/s}$.
}.
This shows that the normal fluid reaches another state with the deformed velocity profile.
In this letter, we call this state ``the deformation state.''
The smaller velocity $\overline{v}_\text{n}=0.6 ~\mathrm{mm/s}$ is unable to trigger a significant deformation of the normal fluid velocity profile.
The point is the onset of the instability, namely the timing when $v_\text{n,c}$ starts to decrease significantly.
The onset time is indicated by the arrows in Fig. \ref{uc_phi.pdf}(a).
When the mean velocity $\overline{v}_\text{ns,a}$ increases, the energy-injection increases to accelerate the instability onset.

This instability is systematically understood by the dimensionless parameter $\phi$ of the mutual friction force of Eq. (\ref{eq:phi}).
Figure \ref{uc_phi.pdf}(b) shows the values of $\phi$ as a function of time, corresponding to Fig. \ref{uc_phi.pdf}(a).
We define the values of $\phi$ at the onset time as its critical value for the velocity deformation.
The points at the onset time are marked by the green circles in Fig. \ref{uc_phi.pdf}(b), and the instability is found to occur for $30 < \phi <70$.
The critical value of $\phi$ for the instability depends on the $\overline{v}_\text{n}$.
According to the linear stability analysis, the Poiseuille profile of the normal fluid becomes unstable when $\phi$ exceeds about 13 \cite{melotte98}.
However, as in Fig. \ref{uc_phi.pdf}(b), the critical values of the deformation are several times larger than that of the linear stability analysis.

The value of $\phi ~(\propto L)$ increases gradually before the onset time, after which it increases rapidly.
This is closely related to the deformation of the normal fluid velocity profile.
As mentioned in Ref. \cite{yui15a}, the Poiseuille normal flow makes the inhomogeneous vortex tangle, and its vortex line density grows at a slower rate than that of the homogeneous vortex tangle with uniform normal flow.
After the onset time, the velocity profile of the normal fluid becomes flatter, which makes the vortex line density grow more rapidly.
In other words, the vortex tangle flattens the velocity profile of the normal fluid such that it accelerates its own growth.

\begin{figure}
  \centering
  \includegraphics[width=1.0\linewidth]{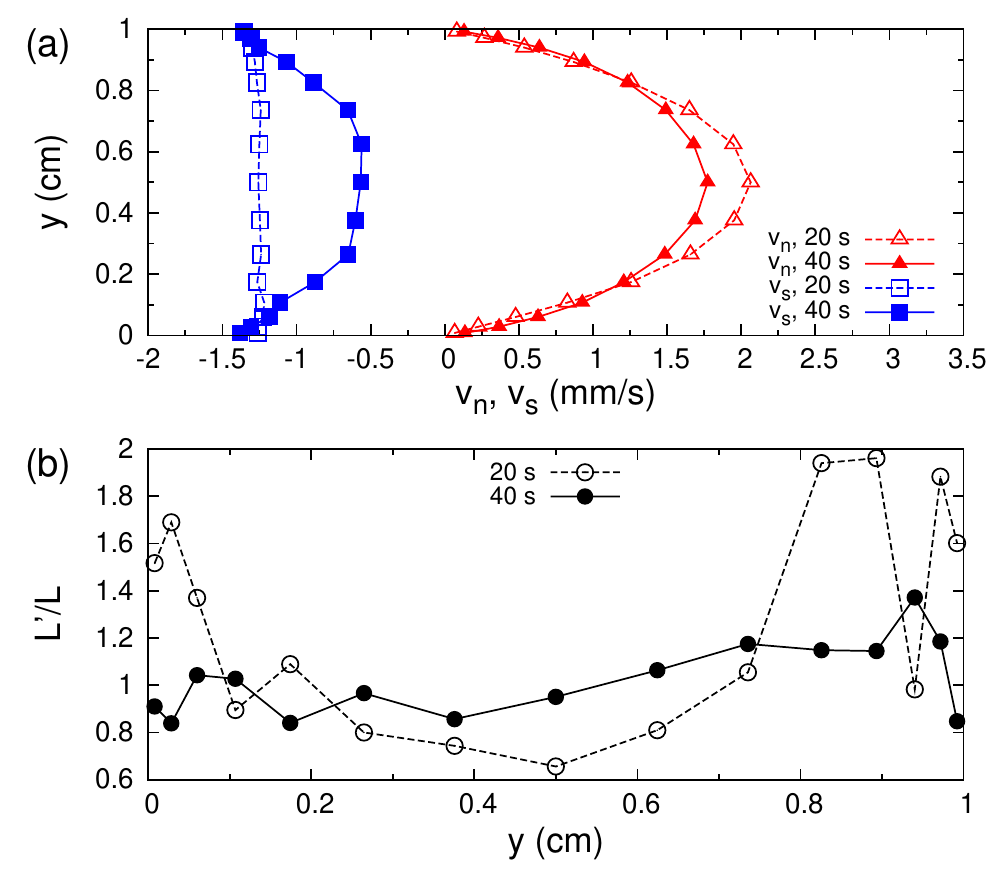}
  \caption
  {
  Simultaneous dynamics of the two fluids: profiles at $z=0.5 ~\mathrm{cm}$ through the center of the channel cross-section at $t=20, 40 ~\mathrm{s}$ for $\overline{v}_\text{n} = 1.0 ~\mathrm{mm/s}$.
  (a)
  The normal fluid velocity $v_\text{n}$ and the superfluid velocity $v_\text{s}$.
  (b)
  The local vortex line density $L'$ scaled by $L$.
  }
  \label{vn_vs_vld_y.pdf}
\end{figure}

In this study, we can investigate the simultaneous dynamics of the two fluids.
It is important to reveal how the two fluids affect each other.
We found that the flattening of the normal fluid velocity is caused by the interaction.
Figure \ref{vn_vs_vld_y.pdf}(a) shows the velocity profiles at $t=20, 40 ~\mathrm{s}$ for $\overline{v}_\text{n} = 1.0 ~\mathrm{mm/s}$.
Here, the superfluid velocity is calculated by ${\bm v}_\text{s} ({\bm x}) = {\bm v}_{\text{s},\omega} ({\bm x}) + {\bm v}_\text{s,b} ({\bm x}) + {\bm v}_\text{s,a} ({\bm x})$.
At $t = 20 ~\mathrm{s}$, the profile $v_\text{n}$ remains nearly parabolic, and the profile $v_\text{s}$ is almost same with the uniform applied velocity $v_\text{s,a}$.
Because the tangle of the quantized vortices does not yet develop fully, the mutual friction is still small as shown in Fig. \ref{uc_phi.pdf}(b).
Hence, the profile $v_\text{n}$ is not modified much, and the velocity $v_{\text{s},\omega}$ induced by the quantized vortices remains much smaller than the applied one $v_\text{s,a}$.
At $t = 40 ~\mathrm{s}$, the profile $v_\text{n}$ is squashed, and the superfluid flow is reduced around the center.
This implies that the relative velocity $v_\text{ns}$ tends to be uniform to decrease the mutual friction, namely the profiles of $v_\text{n}$ and $v_{\text{s},\omega} + v_\text{s,b}$ tend to mimic each other.
Figure \ref{vn_vs_vld_y.pdf}(b) shows the profile of the local vortex line density $L'({\bm x}) = (1/\Omega'({\bm x})) \int_{{\mathcal L}'({\bm x})} \text{d} \xi$.
At $t = 20 ~\mathrm{s}$, the vortices concentrate near the channel walls.
This comes from that the profile $v_\text{ns}$ is spatially nonuniform, and the quantized vortices are nonuniformly affected by the mutual friction: the terms including $\alpha$ and $\alpha'$ in Eq. (\ref{vortex}) are significantly nonuniform.
At $t = 40 ~\mathrm{s}$, the profile $L'$ tends to be uniform, because the profile $v_\text{ns}$ becomes more uniform.
These cooperative dynamics of the two fluids cause the flattened velocity profile of the normal fluid.

\begin{figure}
  \centering
  \includegraphics[width=1.0\linewidth]{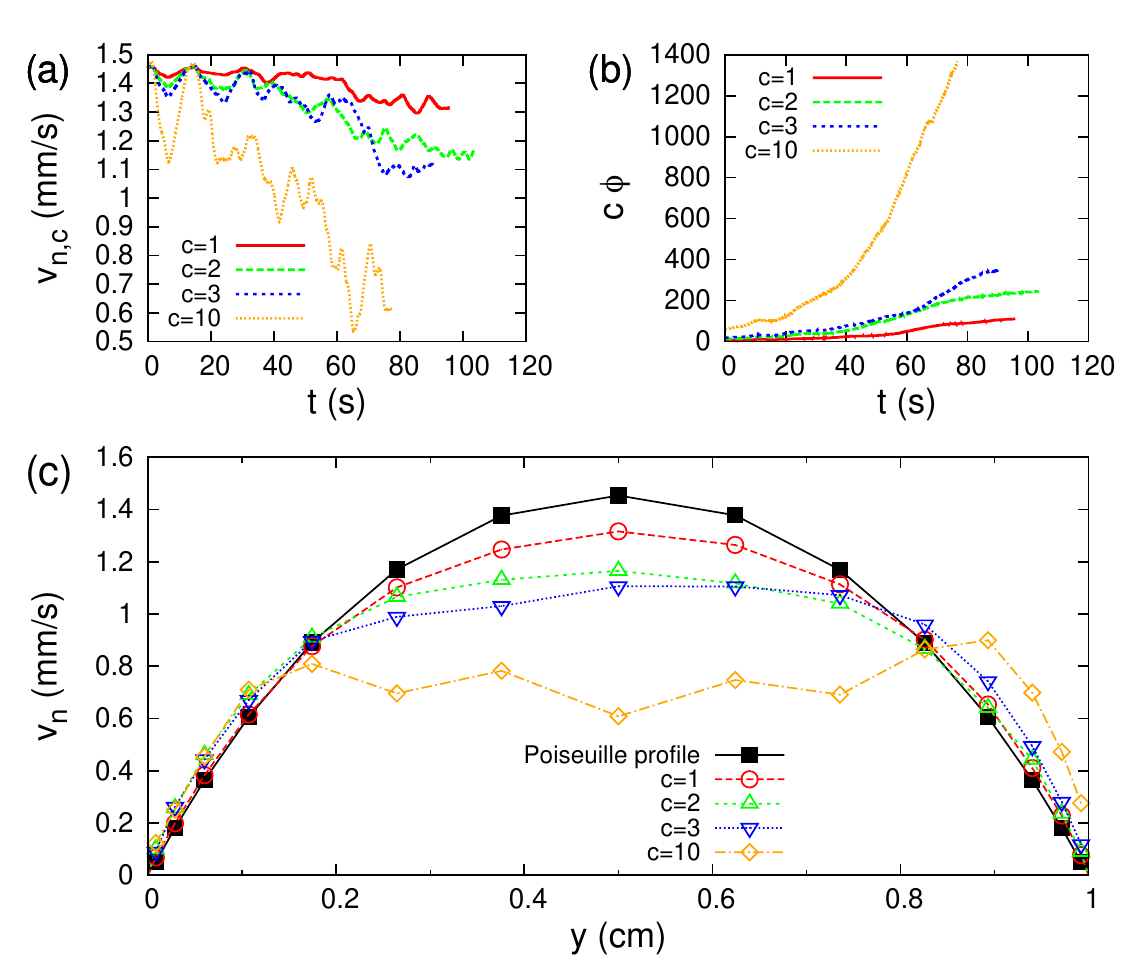}
  \caption
  {
  Results for $\overline{v}_\text{n} = 0.7 ~\mathrm{mm/s}$ with a different scale factor $c$ of the mutual friction force.
  (a)
  The central velocity $v_\text{n,c}$ of a normal fluid as a function of time.
  (b)
  Amplified mutual friction force $c \phi$ as a function of time.
  (c)
  Snapshots of velocity profiles of normal fluid in the deformation state at $z=0.5 ~\mathrm{cm}$ through the center of the cross-section.
  \label{cfns.pdf}
  }
\end{figure}

In order to investigate the velocity profile in the deformation state, we multiply $F_\text{ns}$ by a scale factor $c \geq 1$ in Eq. (\ref{eq:navier}): $F_\text{ns} \rightarrow cF_\text{ns}$.
Here, we keep the values of $\alpha$ and $\alpha'$ in Eq. (\ref{vortex}), which means that the mutual friction from the normal fluid to the superfluid is not amplified.
While this is artificial, it is useful to know how the larger mutual friction deforms the velocity profile.
The numerical simulation is performed for $\overline{v}_\text{n}=0.7 ~\mathrm{mm/s}$ while changing the values of $c$.
As shown in Fig. \ref{cfns.pdf}(a), in each case, the normal fluid reaches the deformation state, where the value of $v_\text{n,c}$ fluctuates around some constant value.
The central velocity decreases more with $c$.
This comes from that the value of the amplified mutual friction $c\phi$ becomes larger with $c$, as shown in Fig. \ref{cfns.pdf}(b).
Figure \ref{cfns.pdf}(c) shows the velocity profiles of normal fluid at $z=0.5 ~\mathrm{cm}$
\footnote
{
The supplemental data are shown in Section IV of the supplemental material.
}.
The velocity profile becomes flatter as $c$ increases.
The profile for $c=10$ is not exactly the same as that of the experiments \cite{marakov15}.
In the experiments, the velocity profile is flattened near the walls but not fully flattened in the central region.
In Fig. \ref{cfns.pdf}(c), the latter behavior does not appear, and the profile close to the walls keeps nearly parabolic unlike the experiments.
In works such as in Ref. \cite{marakov15}, the values of $\phi$ are of the order of $10^3$.
In our simulation, because $c \phi \sim 10^3$ at $c=10$, the experiments may be interpreted as the case of $c=10$ in Fig. \ref{cfns.pdf}(c).
The velocity profile is largely flattened for $c=10$, and this is consistent with the experiments.

{\it Conclusions.}---The coupled dynamics of two fluids has been the frontiers of low temperature physics.
We constructed the numerical method of the coupled dynamics, and performed the numerical simulation of the thermal counterflow in a square channel.
To study systematically the deformation of the normal fluid velocity profile, we introduced the dimensionless parameter.
Using this parameter, we analyzed the extent of the deformation and the critical values for the instability.
The significantly flattened velocity profile of the normal fluid was obtained when the mutual friction force was sufficiently strong.
These results are consistent with the recent visualization experiments.

\begin{acknowledgments}
We would like to acknowledge W. F. Vinen and W. Guo for their useful discussions.
This work was supported by JSPS KAKENHI Grant No. 17K05548 and MEXT KAKENHI ``Fluctuation \& Structure'' Grant No. 16H00807.
S. Y. was supported by Grant-in-Aid for JSPS Fellow Grant No. JP16J10973.
The work of H. K. is supported in part by the MEXT-Supported Program for the Strategic Research Foundation at Private Universities ``Topological Science'' (Grant No. S1511006) and Keio Gijuku Academic Development Funds.
\end{acknowledgments}

\setcounter{equation}{0}
\renewcommand{\theequation}{S.\arabic{equation}}

\section{SUPPLEMENTAL MATERIAL}

\appendix

\subsection{I. Coarse-grained mutual friction force}
Consider the mutual friction force in quantum turbulence on the scale larger or smaller than the inter-vortex spacing of the superfluid.
On a small scale, the superfluid vorticity ${\bm \omega}_\text{s}$ is localized only on the position ${\bm s}$ of the vortex filaments as
\begin{equation}
  \bm{\omega}_\text{s}({\bm x}) = \kappa \int_{\mathcal L} \text{d} \xi {\bm s}' (\xi) \delta({\bm x} - {\bm s}(\xi)),
\end{equation}
where $\delta({\bm x})$ is the Dirac delta function.
Because the mutual friction force is the interaction between the normal fluid and the vortex filaments, it works only on the position ${\bm s}(\xi)$ of the vortex filaments.
On a larger scale, the mutual friction force can be regarded as a coarse-grained averaged quantity over the local sub volume larger than the vortex spacing.
The coarse-grained mutual friction force is
\begin{equation}
  {\bm F}_\text{ns} ({\bm x}) = \frac{1}{\Omega'({\bm x})} \int_{{\mathcal L}' ({\bm x})} {\bm f}(\xi) \text{d}\xi,
  \label{eq:fns}
\end{equation}
where ${\bm f}$ is the mutual friction force per unit length, and is defined as
\begin{equation}
  \frac{{\bm f}(\xi)}{\rho_\text{s} \kappa} = \alpha {\bm s}' \times [ {\bm s}' \times ({\bm v}_\text{n} - \dot{\bm s}_0) ] + \alpha' {\bm s}' \times ({\bm v}_\text{n} - \dot{\bm s}_0).
\end{equation}
Here, $\Omega'({\bm x})$ is the local sub volume at ${\bm x}$, and the integral path ${\mathcal L}'({\bm x})$ is the filament in $\Omega'({\bm x})$.
The Lagrangian coordinate $\xi$, which is advected by the flow, is the arc-length along the vortex filaments, and ${\bm f}(\xi)$ is defined on the filaments.
During the coarse-graining process of Eq. (\ref{eq:fns}), the mutual friction force is spatially averaged, and it is converted to the quantity with Cartesian coordinates ${\bm x}$.
In this study, because the Navier--Stokes equations are solved using the Cartesian coordinates, the expression of Eq. (\ref{eq:fns}) is a suitable formulation.

\begin{figure}
  \centering
  \includegraphics[width=0.8\linewidth]{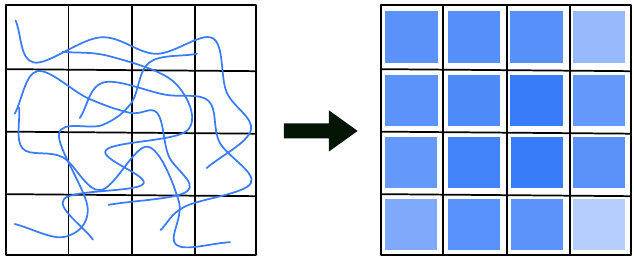}
  \caption
  {
  Schematics of the coarse-graining procedure of the mutual friction.
  The mutual friction on the vortex filaments is averaged over the sub volume.
  The coarse-grained mutual friction is redefined on the sub volumes.
  }
  \label{cg_fns.pdf}
\end{figure}

The treatments of Eq. (\ref{eq:fns}) in the numerical simulation are described below.
The schematics of the coarse-graining procedure are shown in Fig. \ref{cg_fns.pdf}.
The local sub volume $\Omega' ({\bm x})$ corresponds to the cubics divided by the black lines.
The coarse-graining average for ${\bm F}_\text{ns} ({\bm x})$ in Eq. (\ref{eq:fns}) is performed over this local sub volume at ${\bm x}$.
${\bm F}_\text{ns}$ is distributed over the local sub volume as the value defined on the Cartesian grid ${\bm x}$.
In this simulation, the sub volume corresponds to the computational mesh for the normal fluid, and the normal fluid at ${\bm x}$ is acted on by the coarse-grained mutual friction ${\bm F}_\text{ns} ({\bm x})$.

\begin{figure}[b]
  \centering
  \includegraphics[width=1.0\linewidth]{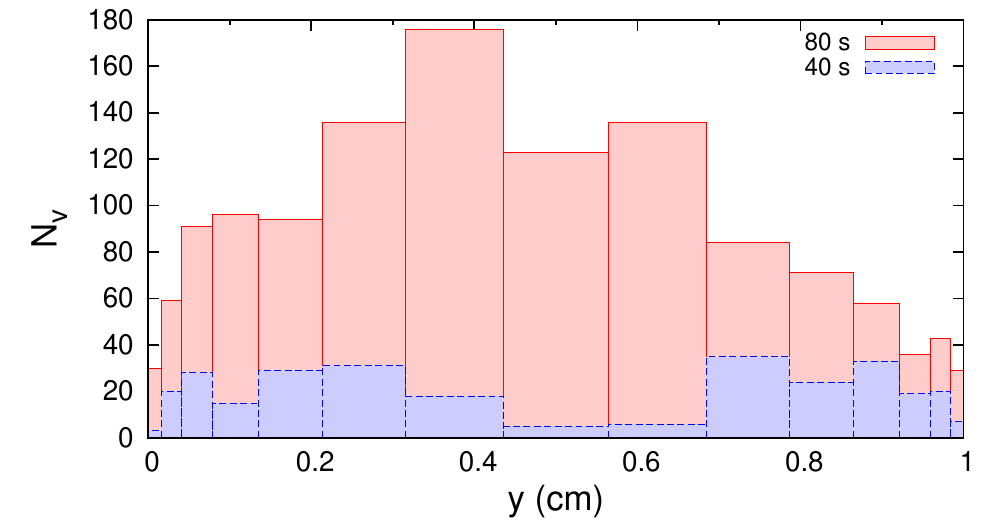}
  \caption
  {
  The number of the discretized points of the vortex filaments at $z=0.5 ~\mathrm{cm}$ for $T=2.0 ~\mathrm{K}$ and $\overline{v}_\text{n} = 0.7 ~\mathrm{mm/s}$.
  }
  \label{points.pdf}
\end{figure}

\begin{figure*}
  \centering
  \includegraphics[width=0.8\linewidth]{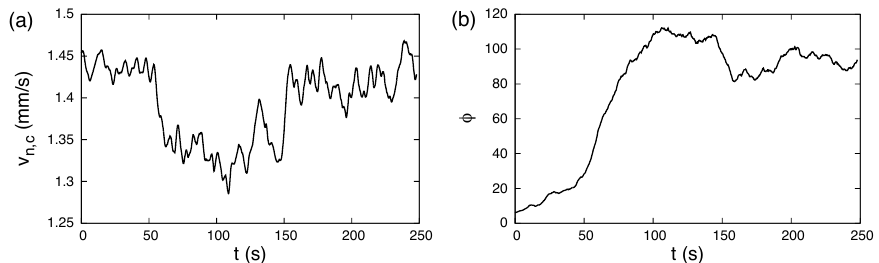}
  \caption
  {
  (a)
  Normal fluid velocity $v_\text{n,c}$ at the center of the channel cross-section as a function of time.
  The parameters are $T=2.0 ~\text{K}$ and $\overline{v}_\text{n} = 0.7 ~\text{mm/s}$.
  (b)
  Mutual friction parameter $\phi$ scaled by viscous force as a function of time.
  }
  \label{uc_phi_20k07vlong.pdf}
\end{figure*}

We check the condition that the coarse-graining procedure of the mutual friction is valid, namely the sub volumes for the coarse-grained mutual friction have more than a few number of the discretized points of the vortex filaments.
In our formulation, because the mutual friction is averaged over the flow direction $x$, the sub volume of the mutual friction is $2D_x \text{d}y \text{d}z$, where $2D_x$ is the length in the $x$ direction of the computational box.
Here, $\text{d}y$ and $\text{d}z$ denote the sub volume widths in the $y$ and $z$ directions, respectively.
We define $N_\text{v}(y_0,z_0)$ as the number of the discretized points in a region
\begin{equation}
\begin{split}
  &0 < x < 2D_x \\
  \text{and}~~ &y_0 - \frac{\text{d}y}{2} < y < y_0 + \frac{\text{d}y}{2} \\
  \text{and}~~ &z_0 - \frac{\text{d}z}{2} <  z < z_0 + \frac{\text{d}z}{2},
\end{split}
\end{equation}
where $y_0$ and $z_0$ are the positions of the discretized points of the normal fluid.
Figure \ref{points.pdf} shows the number $N_\text{v}$ of the discretized points at $T = 2.0 ~\mathrm{K}$ and $\overline{v}_\text{n} = 0.7 ~\mathrm{mm/s}$ in Fig. 2 of the main manuscript.
At $t=40 ~\mathrm{s}$, the value of $N_\text{v}$ is smaller than $10$ very near the channel walls and in the center.
Although the points are dilute in some regions, the mutual friction is small, and the normal fluid is not largely affected by the diluteness.
At $t=80 ~\mathrm{s}$, the value of $N_\text{v}$ becomes large enough to use the coarse-graining procedure.

\subsection{II. Statistically steady state of two fluids}
We performed another longer simulation for $0.7 ~\text{mm/s}$ with a larger time resolution $\Delta t =0.004 ~\text{s}$.
Figure \ref{uc_phi_20k07vlong.pdf}(a) shows the normal fluid velocity $v_\text{n,c}$ at the center of the channel cross-section.
The value of $v_\text{n,c}$ decreases at $t=50 ~\text{s}$, and it briefly fluctuates around some constant value for $50 ~\text{s} < t < 150 ~\text{s}$.
Eventually, it returns to about the initial value at $t=150 ~\text{s}$.
This may indicate that there are three states of the normal fluid, namely (i) the Poiseuille flow with dilute vortices for $0 ~\text{s} < t < 50 ~\text{s}$, (ii) the flattened flow for $50 ~\text{s} < t < 150 ~\text{s}$, and (iii) the parabolic flow with strong mutual friction for $150 ~\text{s} < t$.
State (iii) is different from (i), because the parabolic flow (iii) comes not from the viscous forces but the mutual friction force.
The flattened flow is not maintained in these conditions.
If the mutual friction force becomes sufficiently strong, the flattened flow will be statistically steady.

The state of the superfluid is characterized by the dimensionless mutual friction parameter $\phi$ because it is proportional to the vortex line density $L$.
The value of $\phi$ is shown in Fig. \ref{uc_phi_20k07vlong.pdf}(b) as a function of time.
The value gradually increases for $t < 50 ~\text{s}$, and rapidly increases for $50 ~\text{s} < t < 100 ~\text{s}$.
For the period $50 ~\text{s} < t < 100 ~\text{s}$, the normal fluid velocity profile is flattened, and the development of the vortex line density is accelerated.
For $100 ~\text{s} < t$, the mutual friction saturates, and the value of $\phi$ fluctuates about some constant values.
At $t \sim 150 ~\text{s}$, the normal fluid velocity profile returns to the parabolic profile.
Because of the change of the normal fluid flow, the mean value of $\phi$ changes to another value at $t \sim 150 ~\text{s}$.

Under these conditions, the statistically steady state of the normal fluid is found to be the parabolic flow of the normal fluid.
During the early period, the deformation of the normal fluid velocity occurs with increasing mutual friction.
In the middle period, the normal fluid velocity profile is flattened, whereas the mutual friction force rapidly increases and becomes saturated.
Eventually, the normal fluid velocity has a parabolic profile with a statistically steady quantum turbulence.
The flattened flow will be statistically steady when the mutual friction force becomes much stronger.

\begin{figure*}
  \centering
  \includegraphics[width=0.9\linewidth]{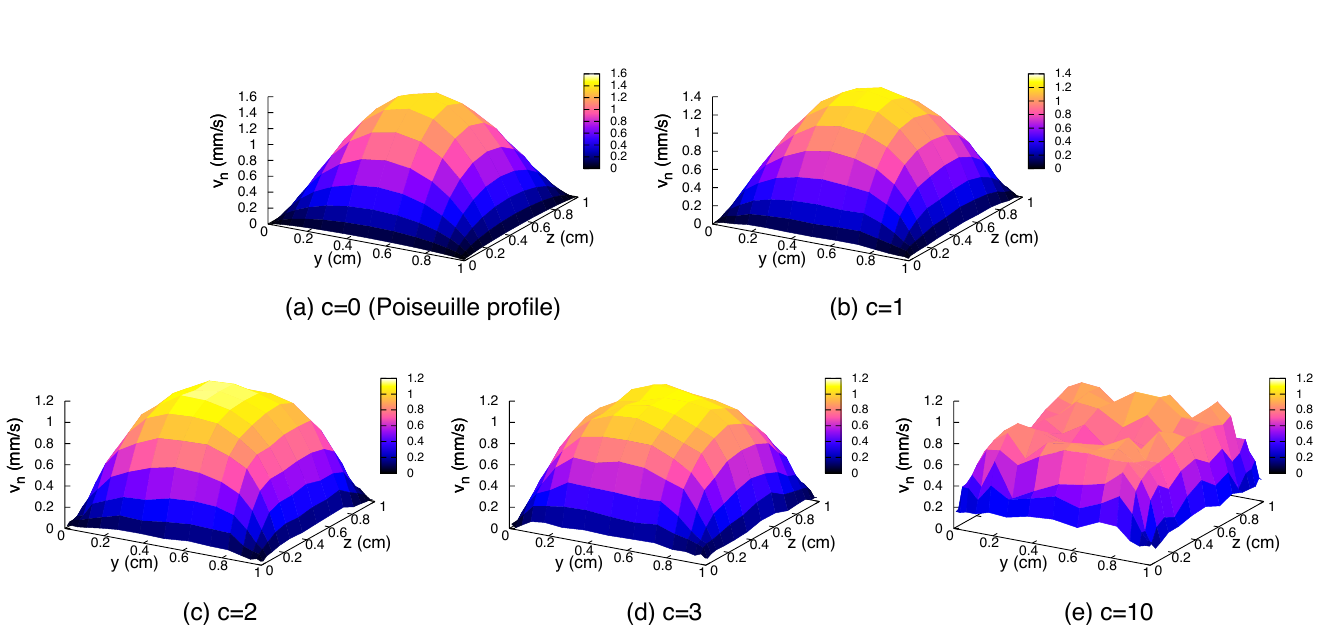}
  \caption
  {
  Velocity profile of the normal fluid over the channel cross-section in the deformation state with the amplified mutual friction $c {\bm F}_\text{ns}$.
  }
  \label{ux2_cfns.pdf}
\end{figure*}

\subsection{III. Analytical solution of Vinen's equation}
Vinen proposed the equation of motion of the vortex line density $L$, namely the vortex line length in a unit volume, in homogeneous thermal counterflow \cite{vinen57c}:
\begin{equation}
  \frac{\text{d}L}{\text{d}t} = \chi_1 \alpha |{\bm v}_\text{ns}| L^{3/2} - \chi_2 \frac{\kappa}{2\pi} L^2.
  \label{Vinen}
\end{equation}
Here, $\chi_1$ and $\chi_2$ are coefficients.
Vinen's equation has the analytical solution
\begin{equation}
  \frac{L^{1/2}(1-\frac{b}{a}L_0^{1/2})}{L_0^{1/2}(1-\frac{b}{a}L^{1/2})}=\exp \left[ \frac{a}{b}\left(\frac{a}{2} t+ \frac{L_0^{1/2}-L^{1/2}}{L^{1/2}L_0^{1/2}}\right)\right],
  \label{solution}
\end{equation}
where $a = \chi_1 \alpha |{\bm v}_\text{ns}|$ and $b = \chi_2 \kappa/2\pi$.
The value of $L$ starts with the initial value $L_0$ and increases to the equilibrium value $L_{\text{eq}}=(a/b)^2$.
According to Vinen's equation, $L$ does not grow up if $L=0$ at $t=0$.
Equation (\ref{solution}) describes the growth of $L$ from some finite initial value $L_0$.
This initial density $L_0$ may be attributable to remnant vortices that survive even in standing superfluid helium.
If we assume $L_{\text{eq}} > L \gg L_0$, the analytical solution is reduced to the solution
\begin{equation}
  \frac{1}{L^{1/2}}=\frac{1}{L_{\rm{eq}}^{1/2}}+\frac{1}{L_0^{1/2}}\exp \left[ -\frac{a}{b}\left(\frac{a}{2} t-\frac{1}{L_0^{1/2}}\right)\right].
\end{equation}
Thus, $L$ grows exponentially to $L_{\text{eq}}$ with the characteristic time
\begin{equation}
  \tau_\text{L}=2b/a^2 = \frac{4\pi}{\chi_2 \kappa L_\text{eq}}.
  \label{eq:tau_L}
\end{equation}
This characteristic time is related to the ``adjustment time'' required for vortices to grow up when the relative velocity is suddenly increased from some initial value to another larger one \cite{vinen57c}.
This estimated value is consistent with that of vortex growth in the previous numerical simulation under the prescribed normal fluid \cite{adachi10,yui15a}.
From the result in Fig. \ref{uc_phi_20k07vlong.pdf}(b), the characteristic time is estimated as $\tau_\text{L} \sim 100 ~\mathrm{s}$ for $\overline{v}_\text{n} = 0.7 ~\mathrm{mm/s}$, which is the time for saturation of $\phi$.
The result $\tau_\text{L} \sim 100 ~\mathrm{s}$ is consistent with its analytical estimate of Eq. (\ref{eq:tau_L}) with the equilibrium value $L_\text{eq} \sim 100 ~\mathrm{cm^{-2}}$.

\subsection{IV. Numerical simulation with amplified mutual friction force}
The flattened velocity profile with the amplified mutual friction is an important result because the profile is consistent with the recent visualization experiments \cite{marakov15}.
Here, we show the supplemental data of Fig. 4 of the main manuscript, namely the velocity profiles over the channel cross-section.
By observing the value of the central velocity, we found that the normal fluid flow reaches the deformation state.
Figure \ref{ux2_cfns.pdf} shows the velocity profiles of the normal fluid over the channel cross-section in the deformation state.
The profile becomes flatter with increasing $c$, i.e., as the mutual friction parameter $\phi$ increases.
The velocity profile is flattened even for $c=2$ and $c=3$, and significantly flattened for $c=10$.
This is consistent with the recent visualization experiments \cite{marakov15}.


\bibliography{book,aps,other}

\end{document}